\documentclass[conference,compsoc]{IEEEtran}
\usepackage{ifpdf}
\ifCLASSOPTIONcompsoc
  \usepackage[nocompress]{cite}
\else
  \usepackage{cite}
\fi
\ifCLASSINFOpdf
  \usepackage[pdftex]{graphicx}
  \DeclareGraphicsExtensions{.pdf,.jpeg,.png}
\else
\fi
\usepackage{amsmath}
\usepackage{amssymb}
\usepackage{stmaryrd}

\usepackage{breqn}

\usepackage{mathtools}
\DeclarePairedDelimiter{\ceil}{\lceil}{\rceil}

\DeclareMathOperator*{\argmax}{arg\,max}

\usepackage{algorithm}
\usepackage{algorithmic}
\usepackage{array}
\ifCLASSOPTIONcompsoc
  \usepackage[caption=false,font=footnotesize,labelfont=sf,textfont=sf]{subfig}
\else
  \usepackage[caption=false,font=footnotesize]{subfig}
\fi
\usepackage{url}
\usepackage{color}

\begin{document}

\title{Overlapping Community Detection by Local Decentralised Vertex-centred Process}

\author{\IEEEauthorblockN{Ma\"el Canu\IEEEauthorrefmark{1}\IEEEauthorrefmark{2},
Marie-Jeanne Lesot\IEEEauthorrefmark{1}\IEEEauthorrefmark{2} and
Adrien Revault d'Allonnes\IEEEauthorrefmark{3}}
%\author{Anonymous}


% compsoc conferences compared to traditional conferences), use this
\IEEEauthorblockA{\IEEEauthorrefmark{1}Sorbonne Universit\'es, UPMC Univ Paris 06, UMR 7606, LIP6, F-75005, Paris,
France\\
E-mail: $\langle$first\_name$\rangle$.$\langle$last\_name$\rangle$@lip6.fr}
\IEEEauthorblockA{\IEEEauthorrefmark{2}CNRS, UMR 7606, LIP6, F-75005, Paris, France}
\IEEEauthorblockA{\IEEEauthorrefmark{3}Universit\'e Paris 8, EA 4383, LIASD, FR-93526, Saint-Denis, France\\
E-mail: allonnes@ai.univ-paris8.fr}
}


\maketitle

\begin{abstract}
This paper focuses on the identification of overlapping communities, allowing nodes to simultaneously belong to several communities, in a decentralised way. To that aim it proposes LOCNeSs, an algorithm specially designed to run in a decentralised environment and to limit propagation, two essential characteristics to be applied in mobile networks. It is based on the exploitation of the preferential attachment mechanism in networks. Experimental results show that LOCNeSs is stable and achieves good overlapping vertex identification.
\end{abstract}


% \fi
\IEEEpeerreviewmaketitle

\section{Introduction}
\label{intro}
Community detection is a central issue in the domain of computational network science \cite{fortunato_community_2009}. Although no consensus exists regarding the definition of a community, it can be intuitively considered as a denser sub-network, an area displaying more connections within itself than with the rest of the network.
%A recurring pattern in many organisational networks, like social networks, it is useful to help understand the structure of the network and the relations between the involved entities.

%In this context, this paper focuses on small decentralised networks relying on peer-to-peer communications between human-carried devices, such as SmartPhone Ad-hoc Networks (SPANs). This kind of Mobile Ad-Hoc Network may be set up when a number of smartphones, carried by individuals, are close enough to each other to establish a fully connected peer-to-peer wireless ad-hoc network.
%
%Exhibiting a de facto social network structure because of the human factor, a community repartition can be found and exploited to get significant improvements in routing efficiency, understood as delivery ratio and delivery cost, for this kind of network. As a matter of fact, flooding routing, generally used in SPANs, is costly in terms of number of transmitted packets, since each node needs to forward numerous packets even when it is not the recipient \cite{hui_identifying_2009}. As a consequence not only does it overload the network bandwidth, but it is costly in terms of computation and, thus, power usage for the smartphones.
%
%However, the very nature of SPANs and other such mobile networks makes it difficult to identify a sufficiently accurate community structure, because the whole network structure may not be known at a given time and, as such, cannot be processed by existing centralised algorithms.

Many community detection methods partition the network into disjoint communities, but it is known that most real-world network communities are \textit{overlapping}, that is, a node can belong to several communities \cite{palla_uncovering_2005,xie_overlapping_2013}. This feature is useful to single out nodes having a special role, being at the interface between two or more communities. These nodes are often, mistakenly, referred to as \textit{overlapping nodes} (in a network) or \textit{overlapping vertices} (in a graph).

The interpretation of multi-membership depends on the context which the considered graph refers to: in a social networking context an overlapping vertex may represent an individual involved in multiple communities, being therefore a useful element to determine relationships between members of these different communities. In a networking context, overlapping nodes can reveal the points where most peering traffic passes, useful to better route inter-networks traffic.

The identification of overlapping vertices is difficult because of this contextual definition. Usually, not all vertices at the interface between two communities can be defined as overlapping, because overlapping vertices have a more specific function than constituting a border. Should there be too many overlapping vertices, i.e. would nearly every community share a majority of its vertices with others, then the meaning of these overlapping vertices, and even of the whole community structure, would be far less legible. On the other side, virtually any network exhibiting a community structure contains vertices that can be viewed as overlapping. Therefore, designing a detection method accurately identifying overlapping vertices is not trivial.

To address these problems, we propose  \textbf{LOCNeSs} (\textbf{L}ocating \textbf{O}verlapping \textbf{C}ommunities in \textbf{Ne}twork \textbf{S}tructure\textbf{s}) a vertex-centred algorithm able to detect communities in a decentralised way, using a local approach and limiting propagation. This algorithm processes an undirected graph and produces a community cover (i.e. a vertex partition) of this graph, encompassing overlapping vertices.

This novel approach allows in particular an implementation in the specific context of decentralised ad-hoc networks (MANET) \cite{hui_distributed_2007}, whose very nature makes it difficult to identify a sufficiently accurate community structure: the whole network may not be known at a given time and, as such, cannot be processed by existing centralised algorithms. Being fully decentralisable and limiting propagation is also best suitable to be implemented in graph analysis Think-Like-A-Vertex (TLAV) frameworks such as Pregel \cite{mccune_thinking_2015}, because vertex data dependency and propagation both generate much inter-process messages, which are known to impact running time performances \cite{mccune_thinking_2015}.
%Vertex data dependency, induced by a centralised method, as well as data propagation, used by some decentralised methods like label propagation, generates much communication between processes and communication between processes is 

The rest of this paper is organised as follows: Section~\ref{works} reviews existing community detection methods; Section~\ref{method} describes the LOCNeSs algorithm in detail; Section~\ref{expes} presents experimental results validating the proposed approach and the paper concludes in Section~\ref{conclusion}.

\section{Related Works}
\label{works}
This section presents existing methods related to community detection in networks, successively considering the non-overlapping and overlapping cases.

\subsection{Community Detection in Graphs}
\label{communities}
There are numerous community detection methods, we propose here to distinguish three main approaches: criterion-based, label propagation and vertex-centric. They can be further distinguished according to their community discovery strategy, local or global: local detection processes a subgraph in order to determine if it contains one (or more) community, where global detection considers the whole graph and tries to delimit the subsets forming a community structure.

Besides the three previously mentionned main approaches, other popular methods include: clustering (local or global),
spectral graph analysis (mostly global),
random walks and Markov processes  (local or global),
statistical inference/probability methods based on generative models (mostly global). See~\cite{fortunato_community_2009,bedi_community_2016} for extensive surveys and references.

\textbf{Criterion-based} methods constitute a very popular family. They are based on an objective function optimisation. This function measures the quality of the graph partition into communities. It can be global, like the popular network modularity \cite{newman_finding_2004}, which has been proved to yield very good results although suffering from major drawbacks such as resolution limit \cite{fortunato_resolution_2007}, or local, for instance \cite{clauset_finding_2005}, which is a local adaptation of modularity. Adjustments of modularity and other criteria have been made to run on distributed graphs~\cite{hui_distributed_2007} but they do not take the overlapping case into~account.

\textbf{Label Propagation} is a family of local methods, among those offering the best results: each vertex is given a unique label and propagates it to all other vertices throughout the graph~\cite{raghavan_near_2007}. After the propagation phase, each vertex retains the most frequently received label for its community. This technique can be implemented in MANET, however the massive use of propagation overfloods the network and is a major drawback \cite{tian_think_2013}.

%\begin{algorithm}[t]
%\caption{Generic frame of a leader/seed-centric approach}
%\label{algo:seed-centric}
%\begin{algorithmic}[1]
%	\REQUIRE $G = (V, E)$, a connected graph
%	\ENSURE $C \subset \mathcal{P}(V)$, communities as sets of vertices
%	
%	\STATE $C \leftarrow \emptyset$
%	\STATE $S \leftarrow$ compute\_seeds($G$), set of leaders/seeds
%	\FOR {$s \in S$}
%		\STATE $c \leftarrow$ expand\_community($s$)
%		\STATE add $c$ in $C$
%	\ENDFOR
%	
%	\STATE merge($C$)
%		
%\end{algorithmic}
%\end{algorithm}
\textbf{Vertex-centric} approaches rely on the principle that some vertices in the network are ``leaders" or ``seeds" and the others are followers \cite{riedy_detecting_2011}. The terms leader will be used throughout the paper to refer to the central vertex role in the general sense related to seed-centric and leader-based approaches. In these methods, communities are formed by gathering followers around leaders, resulting in ego-centred communities (centred around the leader). An example is the \textit{Top-Leaders} approach \cite{rabbany_top_2010}. Although this method is more related to $k$-means clustering (re-allocation of the leader) than truly vertex-oriented, the introduced idea of expanding communities around leaders considering the potential \textit{preference} of a follower vertex (resp. a group of follower vertices) to join a leader vertex has been exploited by several algorithms. \textit{Leader-Follower} \cite{shah_community_2010} assumes that each community is a clique, which is a strong assumption generally not true in real-world networks. \textit{YASCA} \cite{kanawati_yasca:_2014} greedily expands communities around leaders and gathers communities using ensemble clustering; \textit{LICOD} \cite{yakoubi_licod:_2014} starts with a careful selection of leaders before computing ranked community membership for each follower, then adjusting preferences and memberships until stabilisation using strategies borrowed from social choice theories. \textit{EMc} and \textit{PGDc}~\cite{van_laarhoven_local_2016} locally expand around a leader via EM or Projected Gradient Descent algorithm, using conductance to delimit communities. Canu et al. \cite{canu_fast_2015} consider each vertex as a potential leader and use vertex preference measures to constitute dependencies between vertices.%In this case, the community is not ego-centred, one or more leader, the core of the community, are at the top of all the dependencies in this community.

\subsection{Overlapping Detection}
Because multi-membership vertices are frequently encountered in real-world networks, numerous overlapping community detection methods have been proposed, see \cite{xie_overlapping_2013} for a thorough survey. Many are adapted from disjoint methods but original methods have been proposed, including clique percolation or expansion, based on $k$-clique finding in the graph, or game-theoretic based frameworks. Although most of these methods correctly identify communities with respect to the ground truth, they lack precision as for the overlapping vertices identification \cite{xie_overlapping_2013}. First, we distinguish the adapted methods, and then we focus on the agent-based algorithm \textit{iLCD}, more related to the work presented in this paper, and finally we present vertex oriented methods.

\textbf{Adapted Methods} like SLPA \cite{xie_slpa_2011} and COPRA \cite{gregory_finding_2010} are label propagation methods extending the principles of~\cite{raghavan_near_2007}. SLPA defines two roles, \textit{speaker} and \textit{listener}. Each vertex takes turn being the speaker (i.e. spreading its label) while the others are listening, until label stabilisation is reached, which can be arbitrarily long. COPRA introduces a belonging coefficient to extend \cite{raghavan_near_2007} for overlapping communities. Clustering methods and local objective function optimisation, with OSLOM \cite{lancichinetti_finding_2011}, have been adapted as well.

However, many specific ones have also been developed. Agent-based methods, such as \textit{iLCD} \cite{cazabet_simulate_2011}, consider each vertex as an agent and each edge as a bond, a relationship between two agents. Communities are formed through an agent simulation process where each vertex can perform actions such as trying to create or join a new community. The detection relies on each agent doing a local computation, and thus follows the same basic principle as LOCNeSs.

Few \textbf{vertex-centric} overlapping community detection method existing to date. The most popular is \cite{whang_overlapping_2013}. However, this method uses a custom PageRank procedure to expand the leader set, which is not decentralised and generates much propagation, therefore unsuitable for use in the considered context.

\section{Proposed Method: LOCNeSs}
\label{method}
\begin{algorithm}[t]
\caption{LOCNeSs - Step 2}
\label{algo:locness}
\begin{algorithmic}[1]
%	\REQUIRE $v \in V$, a vertex
%	\ENSURE $C(v)$, community assigned to $v$
%	
%	\STATE $S_v \leftarrow k_v$ highest-degree neighbours of $v$
	\REQUIRE $G = (V, E)$ a graph,
	
	$A_v \subset V$ the set of leaders of $v$
	\ENSURE $C \subset \mathcal{P}(V)$ a set of vertex communities in $G$
	
	\STATE {\textit{Step 2.1}}
	\FOR{\textbf{each} $v \in V$}
		%\IF{$|A_v| > 1$}
			\STATE $\widehat{a_v} \leftarrow \argmax_{a \in A_v} d_a$
		%\ELSE
		%	\STATE $\exists! a_v \in A_v ~/~ A_v = \{a_v\}$
		%	\STATE $\widehat{a_v} \leftarrow a_v$
		%\ENDIF
		\STATE \textit{merge}($C(v), C(\widehat{a_v})$) (see Description, step 2.1)
%		\IF{$v$ belongs to several expanded communities}
%		\STATE let $S_v$ be the set of these communities seed
%		\STATE let $\widehat{s_v} \in S_v$ such that $\nexists s \in S_v ~/~ d(s) > d(s_v)$
%		\ENDIF
	\ENDFOR
	\STATE ~
	\STATE {\textit{Step 2.2}}
	\FOR{\textbf{each} $v \in V$}
		\IF{$|A_v| > 1$}
			\FORALL{$a_v \in A_v \setminus \{\widehat{a_v}\}$}
				\STATE $C(a_v) \leftarrow C(a_v) \cup \{v\}$
			\ENDFOR
		\ENDIF
	\ENDFOR
	
\end{algorithmic}
\end{algorithm}

This section presents the proposed algorithm, LOCNeSs, standing for \textbf{L}ocating \textbf{O}verlapping \textbf{C}ommunities in \textbf{Ne}twork \textbf{S}tructure\textbf{s}. First, the extension of the preferential attachment, then the principles of the algorithm itself are described. Finally, some properties of LOCNeSs are discussed: complexity, propagation and stability.

We use the following notations: $G=(V,E)$ is an undirected and unweighted graph, where $V$ is the set of vertices, $E$ the set of edges. $G$ is assumed to be fully connected without isolated vertices nor self-loops. We denote $n = |V|$ and $m = |E|$, $d_v$ the degree and $\Gamma(v)$ the set of neighbours of a vertex $v$. $\bar{d}$ is the average degree in $G$. $\mathcal{C} = \{c_1, ..., c_k\}$ is the set of $k$ communities formed after detection, and $C : V \rightarrow \mathcal{P}(\mathcal{C})$ gives the set of community memberships of a vertex~$v$. If $|C(v)| > 1$, $v$ is an overlapping vertex.

\subsection{Extending Preferential Attachment}
\label{principles}
%As detailed below, the method we propose is based on the preferential attachment principle and relies on a local agglomerative vertex-centred mechanism compatible with decentralised execution, in a SPAN or a TLAV framework.

The preferential attachment theory~\cite{barabasi_emergence_1999} describes a commonly observed formation pattern of complex scale-free networks, especially social networks. It states that a network entity tends to be attracted to other entities it shares many links with, creating denser areas in the network. In other words, during the formation of the network, vertices having higher degrees tend to grab new links more easily, resulting in characteristic network shapes and structures.

Based on this principle and the vertex-centric approach, we propose to consider the community detection task from the perspective of a vertex, trying to answer the question: ``Which of my neighbours am I the most related to?" and then join these neighbours to form a community. We adapt a vertex-centred approach allowing a follower to possibly follow more than one leader, defining an overlapping vertex as a vertex belonging to the communities of all of its leaders.

\subsection{Overview of LOCNeSs}
LOCNeSs is based on a set of leaders and followers formed using a vertex-centred leader selection process (see Section \ref{works}). It is required that this method allows a follower vertex $v$ to be associated with several potential leaders.

The implementation proposed in this paper is based on~\cite{canu_fast_2015}, in which each vertex can be a leader or a follower or both, and performs only local computations in each vertex neighbourhood, thus being fully decentralisable. Each vertex selects which of its neighbours it will follow, resulting in an interdependency structure. A final merging step folds up the interdependencies and a community is formed by each group of vertices following one another.

We add two features, detailed in the next subsection to enable multi-membership identification:

\ 1) a follower can select multiple leaders,

\ 2) the final merging is adapted and divided into two parts.

%Il peut paraitre trivial dans sa conception, cependant d'autres modelisation testees pour la preparation de ce travail, par exemple en fusionnant les communautes de tous les sommets d'un meme leader, montrent qu'il y a trop de fusions opérées, la structure communautaire s'en trouve donc denaturee. Il faut garder en tete le phenomene decrit dans l'introduction \ref{intro} qui est que le caractere recouvrant des communautes d'un graphe doit rester minoritaire. Dans le cas contraire, on arrive vite a une structure dont la quasi-totalite des communautes partagent trop de sommets avec d'autres et deviennent donc inutiles pour caracteriser le graphe et envisager une interpretation.

%La solution trouvee offre donc un equilibre entre la quantite de sommets overlapping dans le resultat et la pertinence de ces sommets, comme on peut le voir dans la Section \ref{expes} relative aux experiences.

\subsection{Algorithm Description}
\label{locness}
This subsection describes the important steps of LOCNeSs. The pseudo-code of step 2 is sketched in Algorithm~\ref{algo:locness}. The set of all leaders is denoted $A \subset V$ and the set of the leaders selected by a vertex $v$, $A_v$. 

\subsubsection*{Step 1 - Formation of the Leader Set}
%We replace the selection of an individual preferred neighbour $a_v$ with the set $A_v$, the set of all of $v$'s neighbours with which the agreement is maximised, and leave aside the condition on minimal agreement. Formally:
%\begin{multline}
%A_v = \{u\in\Gamma(v), \nexists x \in \Gamma(v) ~/~ \\
%{agreement}(v, x) > {agreement}(v, u)\}
%\end{multline}
%
%If there is only one vertex maximising the \textit{agreement}, then the case falls back to the previous version of the algorithm. Otherwise, the random choice of a given~$a_v$ no longer occurs.
%
%Note that, as the graph is fully connected, $\forall v \in V, |A_v| \geq 1$ is guaranteed.
%The $A$ set is formed by any relevant method as the ones cited in Section \ref{works}. It constitutes the set of all the vertices being selected as a leader by at least one vertex of the graph:
%
%$\forall u \in A, \exists v \in V ~/~ u \in A_v$
%
%A selection function $f: V \rightarrow \mathbb{R}^+$ is used by every vertex to select its leaders and form the $A_v$ set. A vertex $v$ applies it to each of its potential leaders and keeps those of maximal value. In the chosen implementation, potential leaders are the neighbours of $v$ and the $f$ function is the \textit{agreement}~\cite{canu_fast_2015}.
A set of \textit{potential leaders} is presumed to exist prior to this step. It can be computed by any relevant vertex-oriented method (see Section \ref{works}).

In this step, a preference function $f_v: V \rightarrow \mathbb{R}^+$ is used to select $v$'s leaders and form the $A_v$ set. It is applied to each vertex and each potential leader. A vertex $v$ keeps in its $A_v$ the leaders it maximises $f_v$ with.

In the proposed implementation, the potential leaders of~$v$ are all of its direct neigbours $\Gamma(v)$, and the $f$ function is the \textit{agreement} \cite{canu_fast_2015}.

\subsubsection*{Step 2.1 - Assignment to Community, Merging}
We propose to adapt the use of the preferential attachment principle (see Subsection \ref{principles}) to the overlapping case. To this end, we define for each $v \in V$ a main leader $\widehat{a_v} \in A_v$. This leader serves as a basis for community merging. In the current version of LOCNeSs, $\widehat{a_v}$ is set to be the maximum degree leader of $v$: $\widehat{a_v} = \argmax_{a \in A_v} d_a$, according to the preferential attachment. If several $a$ are of maximum degree, then one is retained randomly.% We believe that further study could allow to improve the relevance of the chosen $\widehat{a_v}$ vertices.

%, because merging all the $C(a)$, $\forall v \in V, \forall a \in A_v$ would cause too many unnecessary mergers, leading to a incoherent global community structure.
After completion, a merging pass is performed: each vertex $v$ forms its own community, then each vertex community~$C(v)$ is merged with $C(\widehat{a_v})$. 

This way of proceeding can seem convoluted, but among other strategies involving merging tested during the conception of the algorithm, it turned out to be the one that works best. Other solutions, for example having  all the followers merging their communities with those of their leaders produces few large size communities, unrelated with the ground truth. The solution presented here provides a compromise between the quantity and accuracy of overlapping vertices, and cohesive community structure, consistent with the ground truth, as shown in Section \ref{expes}.

\subsubsection*{Step 2.2 - Assignment to Community, Additional Communities}
Once the merges have been performed, if $v$ has been identified as overlapping, then it is added to the community $C(u)$, without fusion, for each remaining vertex $u \in A_v$, resulting in:

$\forall u \in A_v \setminus \{\widehat{a_v}\}, v \in C(u)$.

This last step is the one bringing the overlapping repartition: $v$ becomes a member of each community it has a leader in.

\subsection{Algorithm Properties}
This section discusses LOCNeSs' amount of propagation required in terms of number and size of exchanged messages, its stability and algorithmic complexity.

\subsubsection{Propagation}
As mentionned in Section \ref{communities}, too much propagation is a disadvantage for decentralised methods. In LOCNeSs, each vertex $v$ sends messages to its neighbours ($\bar{d}$ on average), which send $v$ back a reply. The estimated average number of messages sent at each step is $\mathcal{O}(2n\bar{d})$. Thus, the total number of message after simplication is estimated on average to $\mathcal{O}(n\bar{d})$.

%Estimation for step 1 depends on the chosen base method, in our case \cite{canu_fast_2015}.

%During step 1 in \cite{canu_fast_2015}, each vertex $v$ gathers the degree of all of its $d_v$ neighbours, resulting in $2n\bar{d}$ messages of size~1 (counting the request and response), on average. Then it shares the results with the same neighbours resulting again in $2n\bar{d}$ messages of average size $\bar{k}$. During step~2.1, the merging process asks each vertex $v$ its $a_v$ and returns its community assignment, resulting in $2n$ messages of size~1. During step~2.2, each vertex $v$ having $|A_v| > 1$ informs the merging process that it also joins the community of every $C(a), a \in A_v \setminus \{\widehat{a_v}\}$, resulting in $2n(\bar{k}-1)$ messages of size~1. All in all, there are an estimated average of $2n(\bar{d}+~\bar{k})$ messages plus $2n\bar{d}$ messages of size $\bar{k}$ exchanged, $2n (2\bar{d}+~\bar{k})$ in the order of $\mathcal{O}(n\bar{d})$.

In comparison, considering that a typical flooding algorithm requires each vertex $v$ to send its label to every other vertex in the graph, emitting a message that goes through every edge of the graph, produces an order of $\mathcal{O}(nm)$ messages. For example, in a graph of $5,000$ vertices with $\bar{d} = 10$, the total number of exchanged message is about ten times lower for LOCNeSs.

\subsubsection{Stability}
LOCNeSs design makes it  nearly deterministic, the only exception being the case when several $\widehat{a_v}$ can be chosen in step 2, and one is drawn at random. It also benefits from the local approach stability \cite{fortunato_community_2009}, a local variation in the graph structure is less able to cause a major change in the community structure than for a global method. This is illustrated in the experimental results (see Section~\ref{expes}).

\subsubsection{Complexity}
%As stated in \cite{canu_fast_2015}, the complexity of step 1 is $\mathcal{O}(n \times \bar{d}~log~\bar{d})$ where $\bar{d}$ is the average degree of the graph, for example $\frac{m}{n}$ for a balanced random graph.
%
%The complexity of step 2 is $\mathcal{O}(n \times \bar{d}^2)$, and of step 3.1 is $\mathcal{O}(n \times n)$ where the counted operations are the community mergers.
%
%Finally, step 3.2 performs $|A_v \setminus \widehat{v}|$ merging operations for each vertex. Given that $|A_v| = k_v$, we get $\mathcal{O}(n \times \bar{k})$, where $\bar{k}$ is the average $k_v$ over all vertices, and with $k_v$ set to $d_v / 2$ as in Section \ref{expes}, we obtain $\mathcal{O}(n \times \bar{d})$
%
%The overall complexity is thus estimated to be $\mathcal{O}(n \times \bar{d}^2)$. However, in a decentralised implementation where each of the $n$ vertices computes its self-concerning informations, running the calculations simultaneously, we can rewrite these complexities and get an overall complexity of:
%\[
%n \times \mathcal{O}(\bar{d}~log~\bar{d}) + n \times \mathcal{O}(\bar{d}^2) + n \times \mathcal{O}(n) + n \times \mathcal{O}(\bar{d}) = n \times \mathcal{O}(\bar{d}^2)
%\]
Classical complexity study on community detection algorithms rely on strong hypothesis, that are mostly unrealistic for real-world complex networks. Because their structure is usually different from a random graph, the worst case complexity being very pessimistic compared to average performance. As a matter of fact, estimating the potential running time and scalability is hard \cite{fortunato_community_2009}.

In LOCNeSs, each vertex runs operations on at most all of its neighbours. Thus, each vertex $v$'s ``individual vertex program" total complexity can be estimated as $\mathcal{O}(d_v^2)$. So we get $n$ operations running in $\mathcal{O}(\bar{d}^2)$, where $\bar{d}$ is the average degree of the graph. There are $n$ merging operations, whose complexity depends on the community size. Writing $\bar{|c|}$ the average size of a community, we can write the total estimated complexity as $\mathcal{O}(\bar{d}^2 + \bar{|c|})$.

\section{Experimental Study}
\label{expes}
A series of experiments has been conducted to study the validity of LOCNeSs compared to existing algorithms, especially its ability to correctly identify overlapping vertices in a decentralised context, as well as its stability and robustness.

First, the datasets and evaluation criteria used for testing are presented, then the general experimental protocol and finally each experiment alongside its results.

\subsection{Datasets}
Both artificial and real-world graphs are used to test the proposed algorithm. Artificial graphs are generated using the classic LFR benchmark \cite{lancichinetti_benchmark_2008}, which creates graphs with a community structure together with a ground truth, i.e. the community membership assignment for each vertex. Real-world graphs are the classic Zachary's karate club \cite{zachary_information_1977}, and the High School Students network proposed in \cite{xie_overlapping_2013}.

%Though exhibiting common community graph properties, such as decreasing power-law degrees, the LFR benchmark does not output entirely realistic community graphs \cite{orman_towards_2013}. 

\subsubsection*{Artificial Graphs Parameters}
\label{parameters}
We use several sets of parameters for the LFR benchmark generator. Their possible values are the ones used in \cite{xie_overlapping_2013}. We specify which value is used for each experiment in the following subsections.

The \textbf{number of vertices $n$} is tested here for several values up to $n = 10,000$. LOCNeSs has been found to work well on graphs up to $n = 1,000,000$, however we could not compute the NMI nor the Omega Index on million-vertex graphs. The \textbf{mixing parameter $\mu$} is related to the intra- and inter-density of communities, and describes how ``well-knit" the graph is. The higher this value is, the less community detection is easy to perform because boundaries of the dense areas are harder to identify. The \textbf{average degree $\bar{k}$} is set to 10, and the\textbf{ maximum degree $maxk$} is set to 50. \textbf{Community size ranges} are the intervals of possible sizes for a community. Two ranges are used, as in \cite{xie_overlapping_2013}: \textbf{$s$} (small) from 10 to 50 vertices, and \textbf{$b$} (big) from 20 to 100 vertices.

The \textbf{number of overlapping vertices} $O_n$ is usually expressed as a percentage of $n$. We use two values: $O_n = 10\%$ i.e. a minority of overlapping vertices, which is generally easy to capture for community detection methods, and $O_n = 50\%$ i.e. half of the vertices are overlapping, which often results in a dramatic drop in detection performance.
	
Note that $O_n = 50\%$ is an extreme value to evaluate the degeneration of method performances and is never encountered as such in real-world graphs. Indeed, overlapping vertices should remain a minority in the graph (see Section~\ref{intro}).

The \textbf{number of memberships} for overlapping vertices $O_m$ is the exact number of different community memberships for each of the $O_n$ overlapping vertices in the graph. We make this parameter vary from 2 to 8. Note that is it actually uncommon to get the same exact number of memberships for all vertices in a real-world graph.

There are a total of 112 combinations, and we generate 10 graphs for each.

\subsection{Evaluation Criteria}
To measure the accuracy of overlapping vertices identification by LOCNeSs, classic criteria are used, see \cite{xie_overlapping_2013}.

\subsubsection{Similarity between partitions} We use the overlapping variant of the popular Normalized Mutual Information (\textit{NMI}) \cite{fortunato_community_2009}, and \textit{Omega Index} \cite{collins_omega_1988}, a variation of the Adjusted Rand Index (ARI) which gives a score of similarity between two partitions of a set, corrected for chance.

NMI ranges in $[0, 1]$ and Omega Index in $(-1, 1]$. Both reach 1 when the two partitions are totally similar, and 0 when the partitions are totally dissimilar for NMI, or similar to the output of a random partition for Omega Index. Abnormally dissimilar partitions are given a negative Omega Index value.
 
The reason why we use these two criteria is that although they both measure the similarity between two partitions, their results are not always similar nor correlated: NMI uses entropy and thus gives more importance to the fact of having a large fraction of pairs of vertices put in the same communities in the two partitions. Thus, it can be high even if the global community repartition is very different from the ground truth, for example is made of numerous little communities. On the contrary, Omega Index is heavily affected by the global community structure and can easily drop if the number and size of detected communities is not the same as in the ground truth.

\subsubsection{Overlapping Vertex Identification}
NMI and Omega Index both capture the matching of the community structure and the identification of overlapping vertices at the same time, and therefore lack the capability of evaluating the relevance of this identification alone. We therefore use the $F_1$-score ($F$-score) to this end, formulated as a classification task: a vertex is correctly classified if it is categorized as overlapping and if it is the case in the ground truth.

The number of memberships is not considered here, a vertex is defined as overlapping if it follows more than 1 other vertex. Precision and recall are computed  separately and used to compute the $F$-score.

\subsection{Experimental Protocol}
We compare LOCNeSs results to that of: COPRA, Greedy Clique Expander (GCE), iLCD, OSLOM and SLPA described in Section \ref{works}. Being agent-oriented, iLCD is LOCNeSs most direct competitor considering the application context. SLPA and COPRA are local \textit{and} decentralised, so they are also competitors to LOCNeSs, although they use much propagation due to being label propagation methods. OSLOM (criterion-based) and GCE (clique-percolation) are shown to provide more comparison about the performance of overlapping community detection, though they could not be straightforwardly applied to the considered context.

All the tested implementations are those made available by their respective authors, run with the recommended default parameters. For COPRA, the parameter~$v$ to control the maximum number of communities per vertex is set to $\ceil{1.5 \times O_m}$, for SLPA we use the recent GANXiS implementation with the parameter $r$ set to 0.3,value obtained after experimental tuning, and for LOCNeSs, the leader selection step uses the default parameters recommended in~\cite{canu_fast_2015}.

Each algorithm is run 10 times on each graph to deal with its non-determinism. Note that for artificial graphs, these 10 runs are done over each of the 10 different instances generated using the same set of parameters.
%All the presented results are the mean values of all these runs with standard deviation as error bars.

\subsection{Quality of Overlapping Vertex Identification}
\begin{figure}[t]
	\centering
	\includegraphics[width=0.40\textwidth]{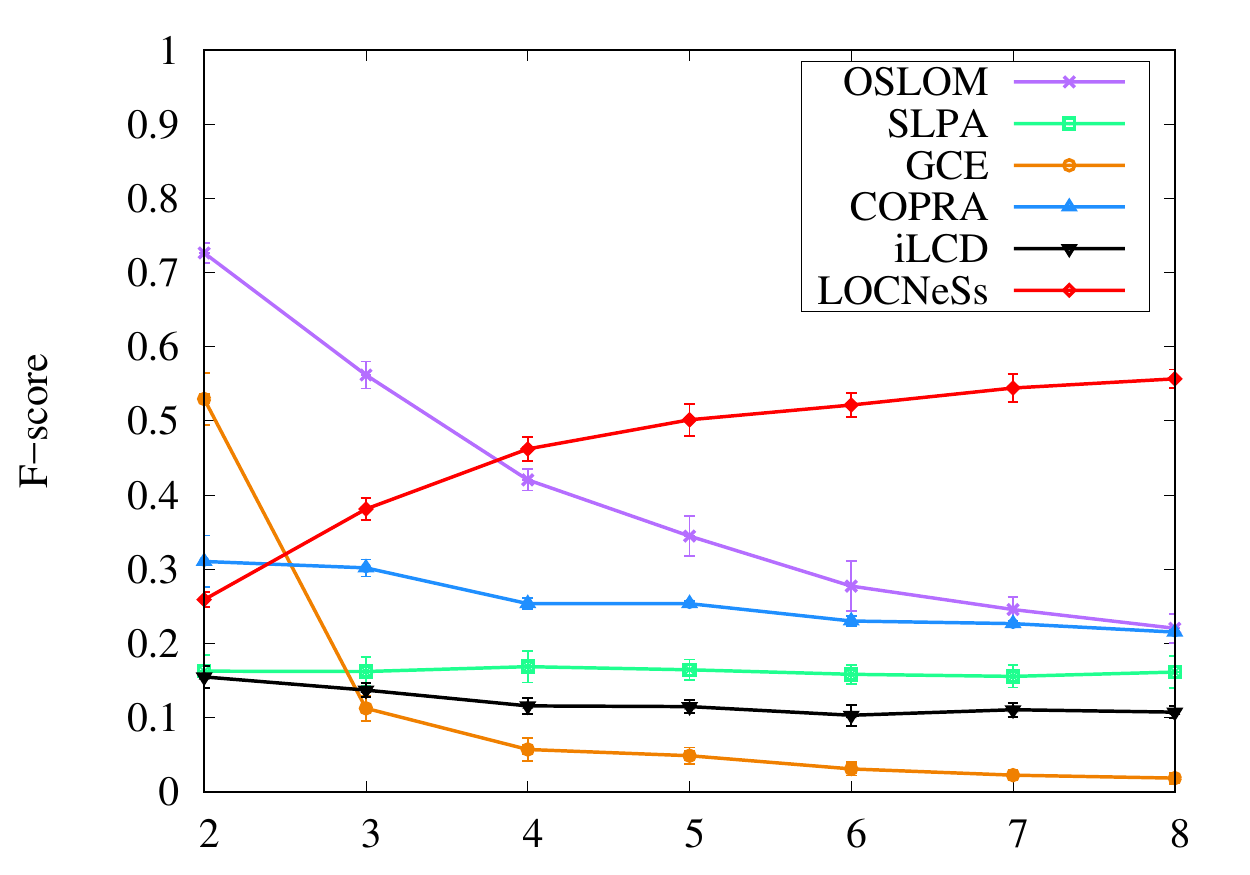}
	\caption{\small Overlapping vertices identification in term of $F$-score as a function of~$O_m$}
	\label{fig:eval_fscore}
\end{figure}

This experiment uses the $F$-score to show the ability of LOCNeSs to identify overlapping vertices, compared to its competitors. This experiment is performed on a set of graphs generated with parameters $n = 5000, \mu = 0.3, O_n = 10\%$, and $s$ community size range. $O_m$ varies.

On the results given Figure \ref{fig:eval_fscore}, we can see that LOCNeSs' $F$-Score rises, contrary to all other methods that decrease or remain stable. In fact, the precision and recall values used to calculate $F$-score, not shown here due to lack of space, reveal that SLPA, GCE and OSLOM are more precise with precision values between 40-50\% for SLPA, but have a lower recall, e.g. around 10\% for SLPA. LOCNeSs achieves only 20\% precision, for $O_m = 2$ to 41\% precision for $O_m = 8$, but has a higher recall: between 34\%, for $O_m = 2$ to 85\%, for $O_m = 8$. In other words, LOCNeSs identifies more overlapping vertices than existing in the ground truth, whereas SLPA identifies less.

Altogether, LOCNeSs achieves significantly better F-score when $O_m \geq 4$, i.e. when a vertex belongs to many communities. This feature is interesting because vertices that are overlapping in certain real world networks, such as social networks, tend to belong to many communities \cite{palla_uncovering_2005}.
% For example on a given graph instance with $O_m = 2$, when the ground truth indicates 500 overlapping vertices, SLPA finds 100 in total of which 47 are correct whereas LOCNeSs finds 870, of which 180 are correct.

\subsection{Quality of Partitions}
\begin{figure*}[t]
	\centering
	\includegraphics[width=0.75\textwidth]{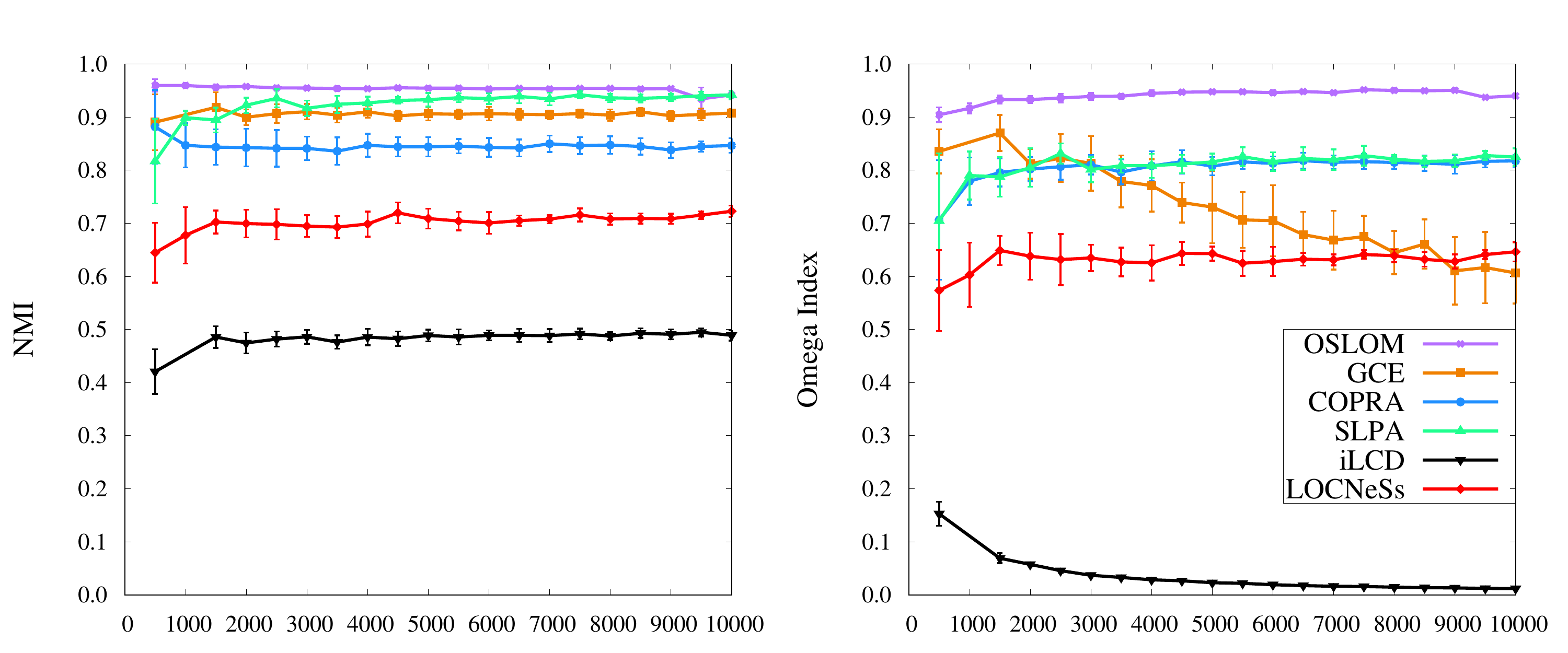}
	\caption{\small Comparative detection quality in terms of NMI and Omega Index as a function of the number of vertices $n$}
	\label{fig:eval_n}
\end{figure*}
\begin{figure*}[t]
	\centering
	\includegraphics[width=0.75\textwidth]{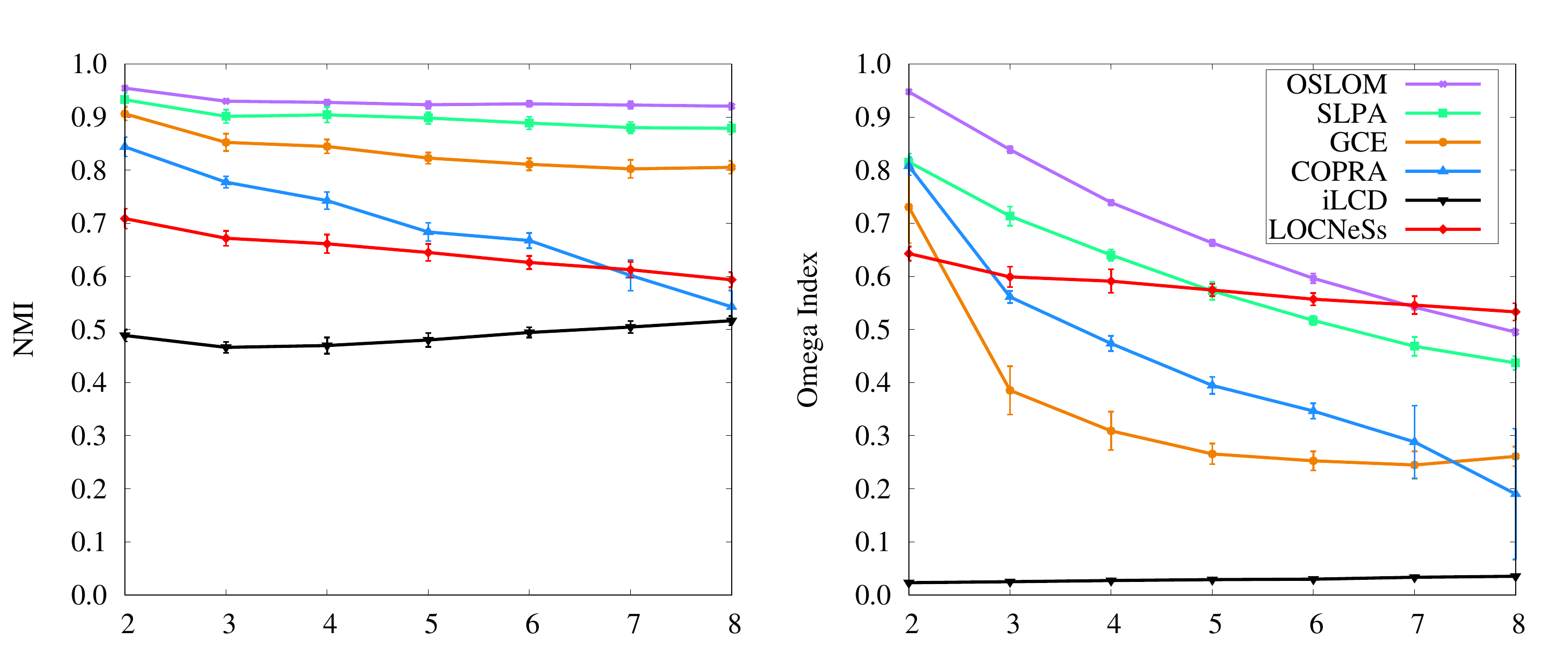}
	\caption{\small Comparative partition quality in terms of NMI and Omega Index as a function of $O_m$}
	\label{fig:eval_nmi}
\end{figure*}
The goal of this experiment is to measure the sensitivity of LOCNeSs against parameters $n$ (Fig. \ref{fig:eval_n}) and $O_m$ (Fig. \ref{fig:eval_nmi}) in the considered graphs. Fixed parameters are $\mu = 0.3, O_n = 10\%$, size range $s$. In Fig. \ref{fig:eval_n}, $n$ varies from $500$ to $10,000$ and $O_m$ is set to 2, and in Fig. \ref{fig:eval_nmi} $O_m$ varies from 2 to 8 and $n$ is set to $5,000$.

We observe that LOCNeSs remains stable in all cases, achieving generally lower performances in terms of NMI and Omega Index than the optimisation or propagation-based methods, but always higher than the agent-based iLCD. In Fig. \ref{fig:eval_nmi} it can be seen that the Omega Index for all other methods drops faster than for LOCNeSs as $O_m$ grows, revealing a consequence of the ability of LOCNeSs to better detect many-memberships overlapping. As a matter of fact, even if the same pairs of vertices tend to be placed together (high NMI), the overall community structure is less and less similar to the ground truth (low Omega Index). That is, having many communities that overlap on the same vertices blurs the community boundaries identification for the other methods.

\subsection{Proportion of Overlapping Vertices and Community Size Range}
\begin{figure*}[t]
	\centering
	\includegraphics[width=0.80\textwidth]{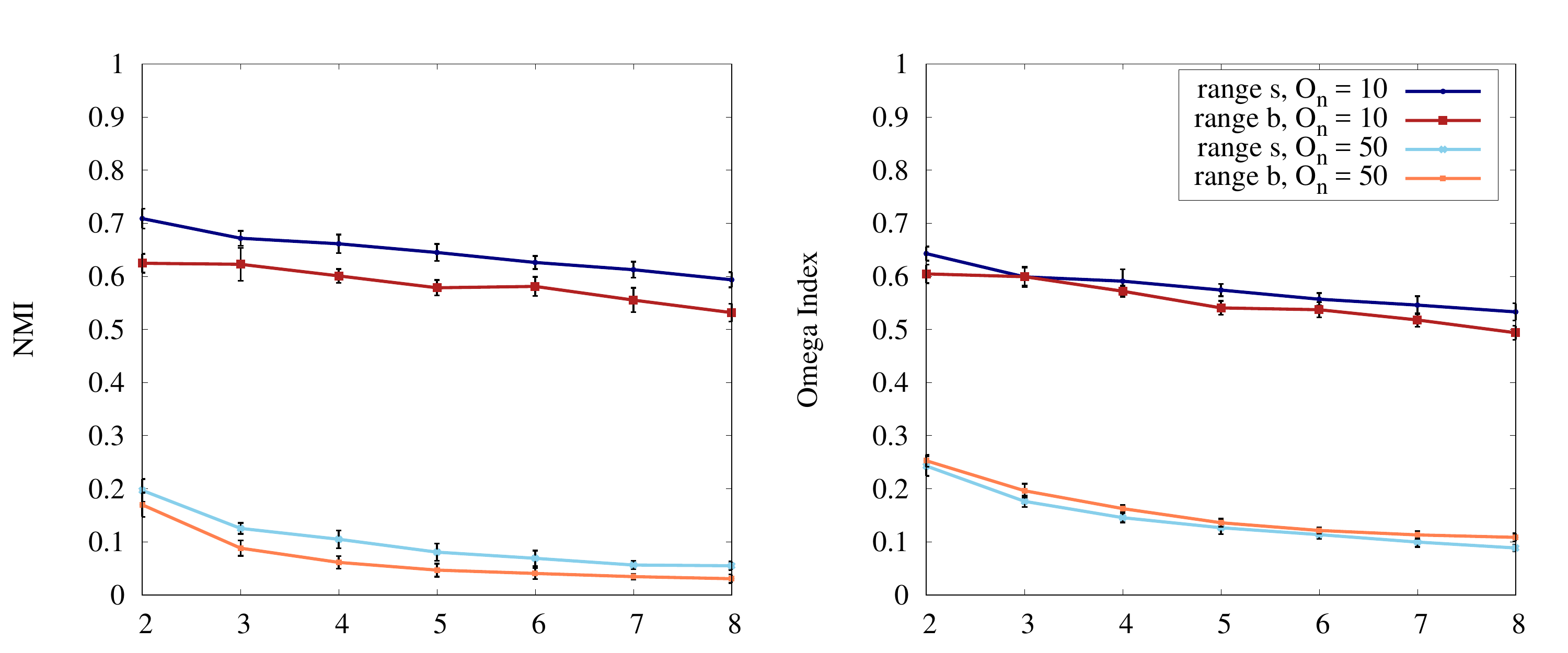}
	\caption{\small $O_n$ value and community size range effect as a function of~$O_m$ for LOCNeSs}
	\label{fig:eval_on}
\end{figure*}

This experiment is intended to measure the impact of two main parameters on the performance of LOCNeSs: the proportion of overlapping vertices $O_n$ and the community size range $s$ or $b$, here shown in term of NMI as a function of $O_m$. Results, presented on Figure \ref{fig:eval_on}, are only related to LOCNeSs due to the lack of space. Similar experiments and results for COPRA, GCE, iLCD, OSLOM and SLPA can be found in \cite{xie_overlapping_2013}. Parameters for the graphs are: $n = 5000, \mu = 0.3$. $O_n$ is tested for two values $10\%$ and $50\%$ while $O_m$ varies. As a reminder, $O_n = 50\%$ is an extreme and unrealistic value, see Section \ref{parameters}.

We notice a significant drop when $O_n = 50\%$ compared to $O_n = 10\%$, which is generally the case for most overlapping community detection methods \cite{xie_overlapping_2013}. NMI for ranges $s$ and $b$ are clearly correlated, performances being higher for the $s$ range, which means that small communities are more easily detected than big ones, a typical feature for methods not relying on modularity, also already identified in \cite{xie_overlapping_2013}.

\subsection{Real-World Graphs}
The last two experiments aim at showing a visual community repartition from LOCNeSs detection results on real-world graphs. The relatively small size of these graphs compared to the artificial graphs makes the NMI and Omega Index values not really significant, therefore we do not include them in the comments. We prefer to highlight some key vertex repartition and try to provide an explanation based on LOCNeSs expected behaviour.
 
\subsubsection{Zachary's Karate Club}
\begin{figure}[t]
	\centering
	\includegraphics[width=0.38\textwidth]{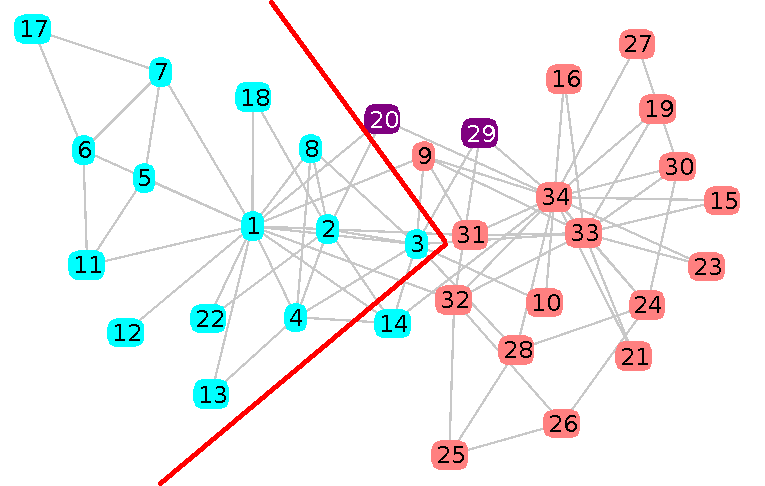}
	\caption{Graph visualisation of the cover produced on Zachary's karate club.
	Orange and cyan: detected by LOCNeSs, including two overlapping in purple.
	Red line: split according to Zachary}
	\label{fig:zachary}
\end{figure}

Zachary's network \cite{zachary_information_1977} pictures a karate club following a clash. Vertices represent the club members and the edges their friendship relations. Note that Zachary's karate club has no ``overlapping" ground truth and was not considered by its author as an overlapping community structure. A visualisation of LOCNeSs detection output is available on Figure \ref{fig:zachary}.
%The visualisation, Figure \ref{fig:zachary}, shows the community structure detected after a single example detection run. Vertex color (orange or cyan) denotes the community assigned by LOCNeSs to this vertex, and the two purple vertices are the ones identified as overlapping. The red separation is the repartition into the two new clubs as mentioned by Zachary in his paper, taken for a ground truth.

We can notice that only a single non-overlapping vertex is misclassified: vertex \#14 has been assigned to the cyan community whereas it is part of the orange one. That is because out of four edges, three link it to the cyan, thus influencing much more the assignment than the only link to the orange to \#34, even if this vertex is of high degree.

Two overlapping vertices are detected: \#$20$ and \#$29$. Vertex \#$20$ shares an edge with \#1, \#2 (cyan community) and  \#34 (orange community), three vertices of high degree in the area, making \#$20$ clearly at the border between the two communities. Vertex \#$29$ is connected to \#$32$ and \#$34$ in the orange community and to \#$3$ in the cyan community, \#$3$ being itself at the border of the community, justifying the overlapping condition of \#$29$ and validating LOCNeSs identification of overlapping vertices.

\subsubsection{High School Network}
\begin{figure}[t]
	\centering
	\includegraphics[width=0.37\textwidth]{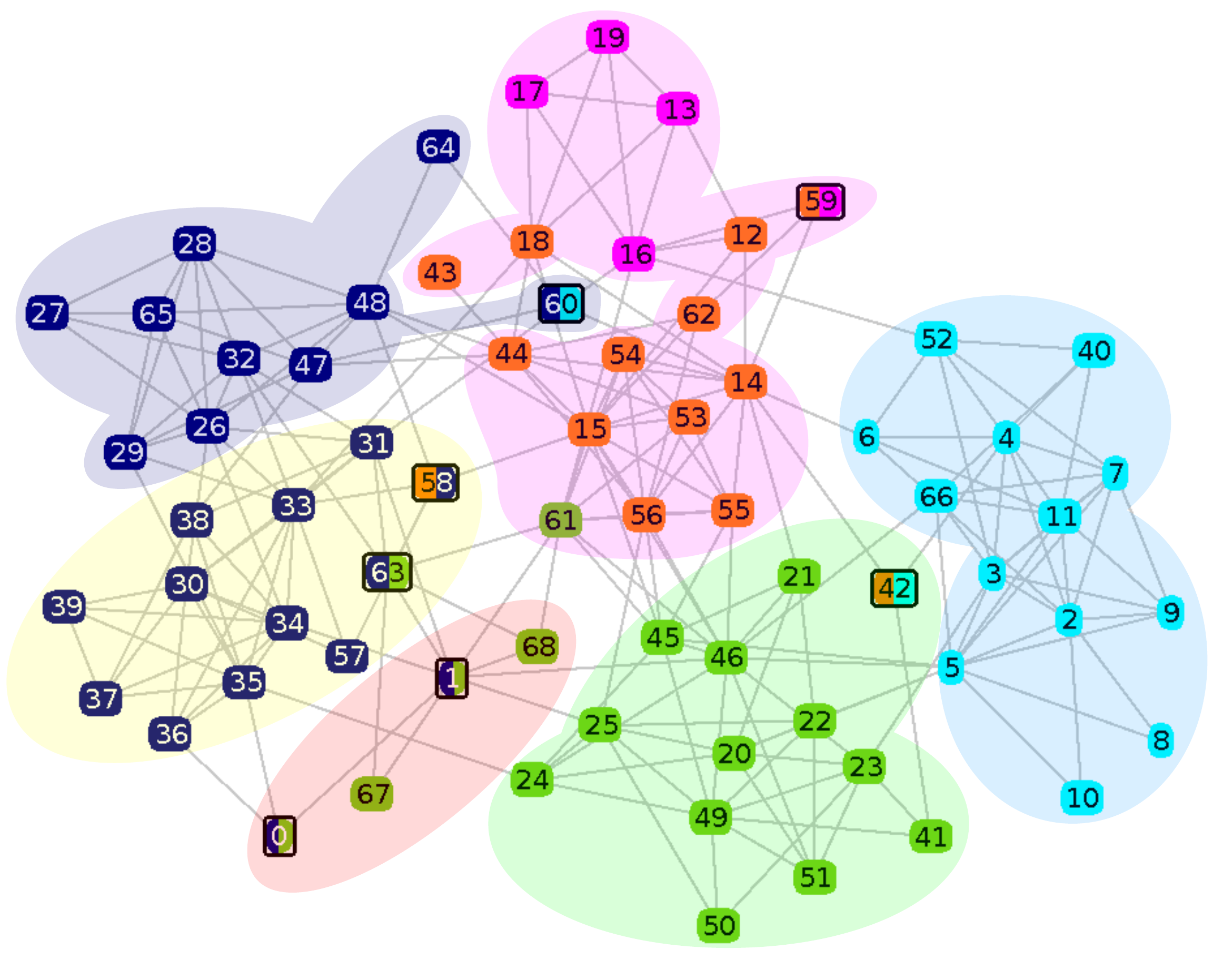}
	\caption{Graph visualisation of the cover produced on the Highschool Network.
	Vertex color: community according to LOCNeSs, background color: ground truth.
	Overlapping vertices are circled in black.}
	\label{fig:highschool}
\end{figure}

The High School Network (Fig. \ref{fig:highschool}) is a real-world network of 68 US high school students, presented by Xie et al. \cite{xie_overlapping_2013}. All the quotations in this subsection are taken from their paper.

This network is a result of a survey of high school students of different grades but from a same school. A link between two students reveals a certain degree of friendship (e.g. frequent interactions). This network encompasses 6 communities corresponding to the students grade, shown as background colors on the figure. A given run of LOCNeSs on the network resulted in 5 different communities detected (vertex color) and 7 overlapping vertices found (\#0, 1, 42, 58, 59, 60, 63), each belonging to two communities, bi-coloured and circled in black on the figure.

We observe that LOCNeSs does not detect the original red community, representing grade 12, but makes vertices \#0 and \#1 belong to both original yellow and green communities, representing respectively grade 11 and 8. This odd proximity between grades 12 and 8 is explained by the misassignment of \#61 to the green community (grade~8) instead of the orange (grade~9). The yellow community (grade~11) is not detected either, being put together with the deep blue one (grade 10).
The preferential attachment disruption between the two areas in the graph is not sufficient to make LOCNeSs decide that these vertices, that form their own community in the real-world, may constitute separate communities. Basically, it corresponds to a lack of density difference.

It is stated in \cite{xie_overlapping_2013} that the original magenta community corresponding to grade 9, here split into the orange and magenta communities, is actually divided into two subgroups: white students (\#14, 15, 41, 43, 44, 53, 54, 55, 56, 59, 61) and black students (\#12, 13, 16, 17, 18, 19). LOCNeSs separated \#13, 16, 17 and 19, in magenta, from the rest of the community in orange rendering, imperfectly, the division. Vertex \#59, assigned to both magenta and orange community, is described as a ``boundary node between subgroups within a grade" .

Lastly, \#42 is described as ``serv[ing] as a bridge between groups without having particular coherence to any group" \cite{xie_overlapping_2013}. This characteristic is well rendered by LOCNeSs that makes \#42 belong to both light blue and green communities.

%The NMI value for this run compared to the ground truth (true division after the grades) is 94\%, which allows LOCNeSs to be ranked top detecting methods for this graph, according to Xie et al. ranking \cite{xie_overlapping_2013}.

\section{Conclusion and Future Work}
\label{conclusion}
We introduced LOCNeSs, an algorithm to detect overlapping communities in graphs. A vertex-centred method, this algorithm implementation is meant to be decentralised and to limit message propagation in the network. These characteristics are particularly suitable for use in a mobile opportunistic network, or to be implemented in a Think-Like-A-Vertex framework, such as Pregel.

We presented experiments to show the efficiency of LOCNeSs, in terms of quality of detection
%with regards to the decentralisation constraint
and stability measured on benchmark graphs, and also in terms of meaningfulness of communities when run on real social networks.

To extend this work, an interesting avenue is to use gradation degree to express the multi-community membership. As a matter of fact, an overlapping vertex can belong more to a particular community than another, which is not reflected with the binary assignment (full belonging or not at all) used here. We are also currently working on a further extension to process dynamic graphs. The experimentally observed robustness against graph topology variations encourages us to think that LOCNeSs would perform well on such time-varying graphs.

\section*{Acknowledgments}
This work was performed as part of the Homo Textilus project, supported by the French National Agency for Research (ANR) under the grant ANR-11-SOIN-007. We thank Chi Dan Pham for helping us to name the algorithm.
%\subsubsection*{Acknowledgments}
%This work was performed as part of the Homo Textilus project, supported by the French ANR agency under the grant ANR-11-SOIN-007. We thank Chi Dan Pham for helping us finding the name of the algorithm.



\bibliographystyle{IEEEtran}
\bibliography{IEEEabrv,bib_pkdd}
\end{document}